\documentclass[%
 aip,
 amsmath,amssymb,
preprint,%
]{revtex4-1}

\usepackage{graphicx}
\usepackage{dcolumn}
\usepackage{bm}

\usepackage[utf8]{inputenc}
\usepackage[T1]{fontenc}
\usepackage{mathptmx}
\usepackage{etoolbox}

\usepackage[colorlinks=true,linkcolor=blue]{hyperref}

\makeatletter
\def\@email#1#2{%
 \endgroup
 \patchcmd{\titleblock@produce}
  {\frontmatter@RRAPformat}
  {\frontmatter@RRAPformat{\produce@RRAP{*#1\href{mailto:#2}{#2}}}\frontmatter@RRAPformat}
  {}{}
}%
\makeatother
\begin{document}


\title[]{Time-resolved force microscopy using delay-time modulation method}
\author{Hiroyuki Mogi}
\affiliation{ 
Faculty of pure and applied sciences, University of Tsukuba, Tsukuba, Ibaraki 305-8573, Japan
}%

\author{Rin Wakabayashi}
\affiliation{ 
Faculty of pure and applied sciences, University of Tsukuba, Tsukuba, Ibaraki 305-8573, Japan
}%

\author{Shoji Yoshida}
\affiliation{ 
Faculty of pure and applied sciences, University of Tsukuba, Tsukuba, Ibaraki 305-8573, Japan
}%

\author{Yusuke Arashida}
\affiliation{ 
Faculty of pure and applied sciences, University of Tsukuba, Tsukuba, Ibaraki 305-8573, Japan
}%

\author{Atsushi Taninaka}
\affiliation{ 
Faculty of pure and applied sciences, University of Tsukuba, Tsukuba, Ibaraki 305-8573, Japan
}%
\affiliation{%
TAKANO Co., Ltd, Miyada-mura, Kamiina-gun, Nagano 399-4301, Japan
}%

\author{Katsuya Iwaya}
\affiliation{%
UNISOKU Co., Ltd., Hirakata, Osaka 573-0131, Japan
}%

\author{Takeshi Miura}
\affiliation{%
UNISOKU Co., Ltd., Hirakata, Osaka 573-0131, Japan
}%

\author{Osamu Takeuchi}
\affiliation{ 
Faculty of pure and applied sciences, University of Tsukuba, Tsukuba, Ibaraki 305-8573, Japan
}%

\author{Hidemi Shigekawa*}
\email{hidemi@ims.tsukuba.ac.jp}
\affiliation{ 
Faculty of pure and applied sciences, University of Tsukuba, Tsukuba, Ibaraki 305-8573, Japan
}%

\date{\today}

\begin{abstract}
We developed a time-resolved force microscopy technique by integrating atomic force microscopy using a tuning-fork-type cantilever with the delay time modulation method for optical pump-probe light. We successfully measured the dynamics of surface recombination and diffusion of photoexcited carriers in bulk WSe$_2$, which is challenging owing to the effect of tunneling current in time-resolved scanning tunneling microscopy. The obtained results were comprehensively explained with the model based on the dipole-dipole interaction induced by photo illumination.
\end{abstract}

\maketitle

Research and development aiming to develop new functionalities by rapidly controlling quantum processes in materials and devices is flourishing, becoming one of the critical challenges. In fundamental properties such as phase transitions, the fluctuations in the dynamics of non-equilibrium systems at the nanoscale define their characteristics. Efforts are also being made in semiconductors to control regions to be as small as $\sim 1$~nm. To achieve these objectives, it is essential to properly understand and utilize the elementary processes in non-equilibrium states. For instance, it is indispensable to elucidate the ultrafast dynamics of carriers and phase changes induced by external fields with high temporal and spatial resolution.

The scanning tunneling microscopy (STM) system can obtain various information in real space with atomic-level spatial resolution, but its time resolution is constrained by factors such as circuitry. On the other hand, with the advancement in quantum optical techniques, pump-probe methods using ultrashort pulse lasers have achieved time-resolved measurements using pulse widths in the femtosecond and attosecond domains. However, typically, there are limitations in spatial resolution. Therefore, over the past decade, efforts have been made to combine these cutting-edge technologies, enabling the development of measurement and control techniques for local spectroscopy with optical time resolution while observing atomic structures and electronic states with atomic-level spatial resolution. Several time-resolved microscopy methods have been realized\cite{Terada, Yoshida_NatNanotech, Yoshida_ACSPhoton, Mogi_NPJ, Arashida_ACSPhoton, Wang, Plankl, Jelic, Yoshioka, Abdo} (see Note 1 in Supplementary Information for more details concerning time-resolved STM).

Time-resolved STM is a highly promising method, but owing to the use of tunneling current in STM, the samples are limited to those with conductivity. Therefore, as a complementary measurement method, the development of time-resolved atomic force microscopy (AFM), which is one of the probe microscopy methods and measures the force between the probe tip and the sample, has been attempted\cite{Hamers, Jahng_APL, Schumacher_PNAS, Araki, Schumacher_APL, Kuroiwa}. For example, interactions based on surface photovoltage (SPV)\cite{Hamers} and dipole induced in probe tip by light (PiFM)\cite{Rajapaksa, Yamanishi}, have been used as a time-resolved force probe\cite{Hamers, Jahng_APL, Schumacher_PNAS, Araki}. Initially, methods synchronizing the cantilever vibration with pulse light excitation were used; however, they had limited time resolution\cite{Araki}. Subsequently, picosecond-range time resolution was achieved, but the S/N ratio was not particularly high\cite{Schumacher_APL}. Other innovations include techniques that use the difference in modulation frequency between pump light and pulse light to remove noise effects\cite{Jahng_APL} and self-detection tuning-fork-type cantilever driven by a quartz oscillator to eliminate the effects of light irradiation used for force measurement\cite{Jahng}. However, these methods all relied on the intensity modulation of the excitation light, similar to the macroscopic optical pump-probe method. 

Here, we demonstrate a time-resolved force microscopy technique developed by integrating a tuning-fork-type frequency modulation (FM)-AFM with the delay time modulation method. We successfully measured the dynamics of surface recombination and diffusion of photoexcited carriers in bulk WSe$_2$, which are challenging in time-resolved STM owing to the effect of tunneling current.

Figure~1(a) shows an overview of the system developed. In this study, we used an optical system with the delay time modulation of two pulse lasers (532~nm, 45~ps pulse width, 500~kHz) that can be driven by an external voltage pulse generator\cite{Iwaya, Mogi_APEX1}. The delay time modulation shown in Fig.~ 1(b) is a method by which the delay time between the pump light and pulse light periodically changes (in this case at 20~Hz) between the measurement time $t_{\rm d}$ and the sufficiently separated time $t_{\rm d, ref}$, and the corresponding signal is detected by lock-in detection. With the delay time modulation method, it is possible to not only eliminate the effects of thermal expansion but also extract only the force component that reflects the dynamics of the sample excited by the pump light\cite{Terada, Mogi_NPJ, Iwaya, Mogi_APEX1, Yoshida_APEX}.

As shown in Fig.~1, a tuning-fork-type cantilever was used, eliminating the need for an optical system for probe vibration detection. This choice makes the method implementable even within complex devices inside a UHV chamber. The spring constant can be high, allowing for a reduced vibration amplitude. The probe tip was attached at an angle to allow for light irradiation directly from the direction of the sample surface. By choosing the probe tip type, both STM and AFM measurements are possible. A tungsten (W) probe, electropolished to a length of $\sim 1$~mm, was attached to the cantilever using silver epoxy paste (The frequency resonance Q-factor $\sim 10000$ was confirmed in Fig.~1(c)). Laser light was introduced through an objective lens (N.A. $\sim 0.4$) positioned right at the top (below) of the probe tip, allowing for light irradiation with a spot diameter of several $\mu$m without interference with the probe structure (Fig.~1(d)). With a 550~nm short-pass filter, the laser light and the light from the lamp for illumination were combined. The position of the spot was observed with a CCD camera placed on the objective lens. For time-resolved force measurement, the amplitude of the frequency shift $\Delta f$ responding to the delay time modulation was determined as a time-resolved signal using a lock-in amplifier.

\begin{figure*}
\includegraphics[width=15cm]{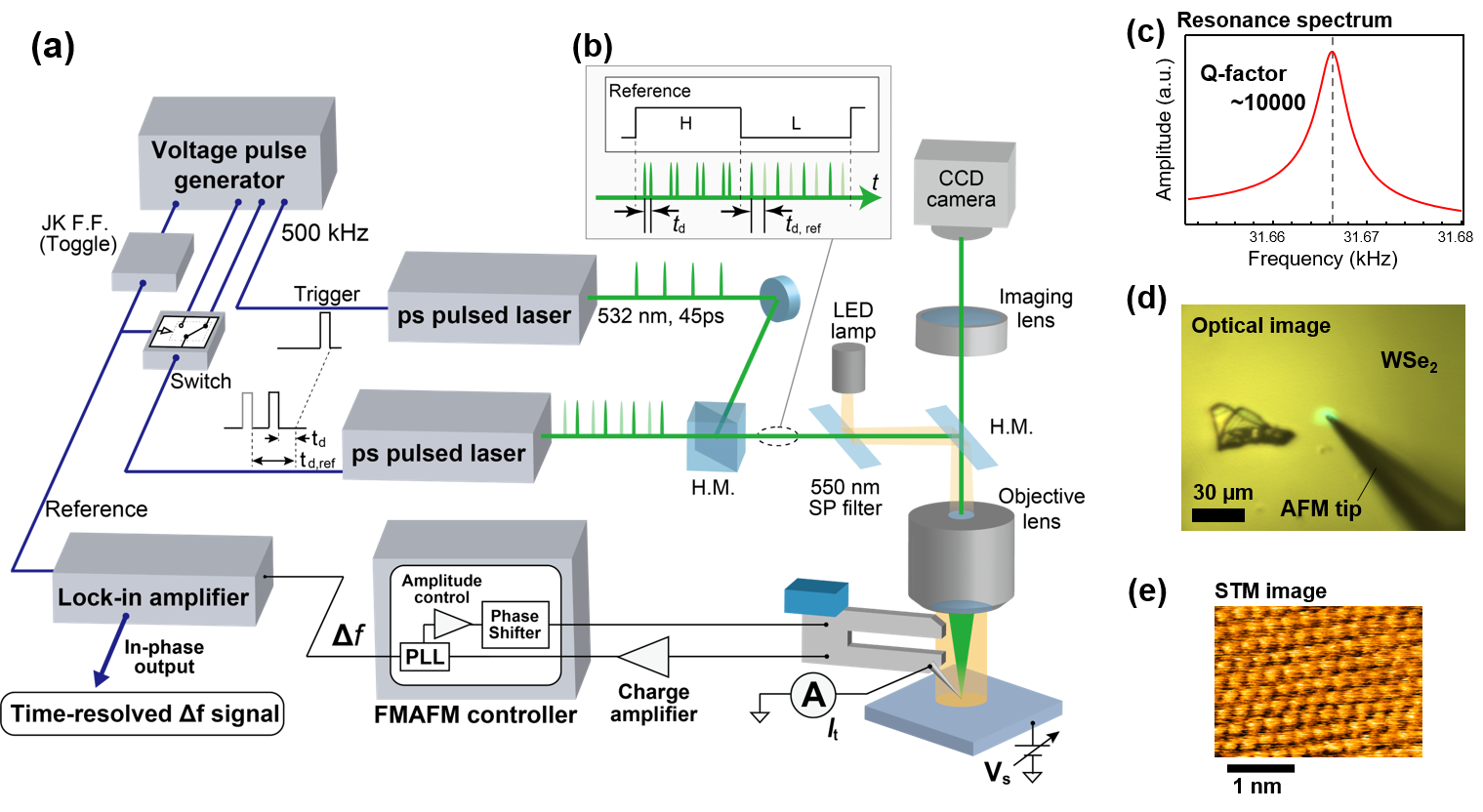}
\caption{(a) Schematic of the time-resolved FM-AFM system developed. H.M.: Half mirror, JK F. F.: JK flip flop, SP filter: short pass filter. Delay time modulation of the two pulse lasers can be driven by an external voltage pulse generator. Inside a UHV chamber ($5\times10^{-8}$~Pa), a tuning fork (fundamental frequency of $\sim 32.768$~kHz) was mounted on an XYZ piezo scanner. The probe tip was positioned at an angle of about 45° to the sample surface. A thin gold-plated W wire was used to electrically connect the probe tip holder and the W tip to measure the current flowing through the probe. (b) Delay time modulation scheme. By adjusting the timing of H/L voltage sequence from JK F.F., the delay time modulation between (H, ${\rm delay\; time} = t_{\rm d}$) and (L, ${\rm delay\;time} = t_{\rm d, ref}$), as shown in the figure, occurs. (c) Resonance characteristics measured to determine the Q factor ($\sim 10,000$). The excitation was achieved by applying an A.C. voltage to the electrodes of the tuning fork, and the vibration signal was detected by a charge amplifier. (d) CCD image of measurement setup. The WSe$_2$ flake on the left side was used to adjust the spot light position. (e) Atomically resolved STM image of a bulk WSe$_2$ sample.
}
\end{figure*}

As the sample, we used WSe$_2$, one of the layered materials (Fig.~1(e)). In previous study using time-resolved STM, photoexcited carrier dynamics, including photoinduced SPV, recombination, and diffusion, were investigated\cite{Yoshida_APEX, Mogi_APEX1}. Considering that the results of our time-resolved AFM can be compared with the findings of the previous study, we conducted experiments on multilayered WSe$_2$.

First, to investigate the band alignment and photoexcited carriers near the surface of the bulk WSe$_2$ sample, we conducted force spectroscopy ($\Delta f$-$V_{\rm s}$ curves, $V_{\rm s}$: sample bias)\cite{Arai} under both dark and illuminated conditions (Fig.~2(a)). We selected three representative conditions: (i) dark state, (ii) weak excitation: only illuminating with the continuous LED lamp used for microscopy (28~$\mu$W; spot diameter, 200~$\mu$m; wavelength, $550 \sim 750$~nm; namely, excitation density of $\sim 90$~pW/${\rm \mu m}^2$), and (iii) strong excitation: in addition to (ii), we illuminated with a pulse laser (1~mW; spot diameter, $\sim 5$~$\mu$m; central wavelength, 532~nm; implying excitation density of $\sim 50$~${\rm \mu W/\mu m^2}$). The averaged results from 20 curves obtained for each condition are shown in Fig.~2(a). The feedback was conducted in the repulsive regime.

By fitting the obtained $\Delta f$-$V_{\rm s}$ curves with a quadratic function for each, we derived the contact potential difference ($V_{\rm cpd}$), under the three conditions: (i) $-0.13 \pm 0.02$~V, (ii) $0.37 \pm 0.01$~V, and iii) $0.92 \pm 0.02$~V, indicating the existence of SPV. SPV is the difference in $V_{\rm cpd}$ with and without illumination, producing the force between the probe tip and the sample. The quadratic curvature became pronounced as the light intensity increased, and $\Delta f$ at the peak tended to increase. The curvature depends on the distance between the probe tip and the sample\cite{Arai}. These results suggest that, owing to the attractive interaction produced upon illumination, the probe tip-sample distance decreased to counteract this attraction owing to the feedback ($V_{\rm s} = -3$~V and $\Delta f = +1$~Hz), increasing the repulsive force at the peak. 

The $I$-$V$ curve obtained by simultaneously capturing the current between the probe tip and the sample $I_{\rm t}$ while measuring the $\Delta f$-$V_{\rm s}$ curve is shown in Fig.~2(b). In the (i) dark and (ii) weak excitation states, almost no current flowed. In contrast, in the (iii) strong excitation state, we detected a current on the order of several hundred fA. This result corresponds to the observation on $V_{\rm cpd}$, where the distance between the probe tip and the sample decreased, thinning the tunneling barrier. The minor current observed within the bandgap region was not noted in previously recorded results of time-resolved STM\cite{Yoshida_APEX}. Namely, for the present sample, it is considered that defects or impurities, which can induce states within the gap, exist at the surface\cite{Addou}. 

\begin{figure}
\includegraphics[width=10cm]{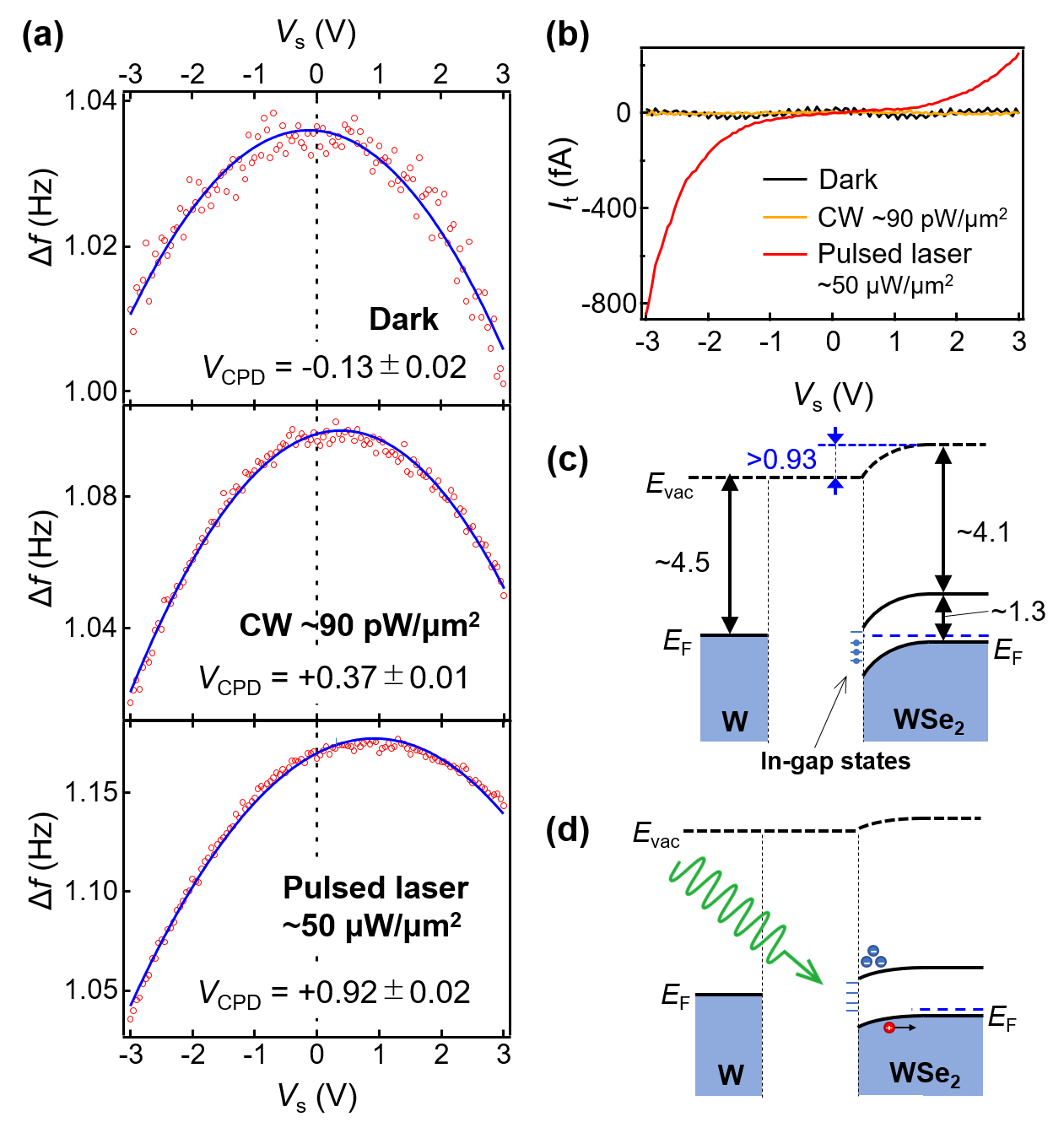}
\caption{
(a) Force spectroscopy results obtained under three representative conditions. $\Delta f$ is the shift of the resonance frequency of cantilever sensitive to the interaction between the probe tip and the sample. After setting the tip position with a repulsive force condition ($V_{\rm s} = -3$~V and $\Delta f = +1$~Hz), we turned off the feedback control and swept $V_{\rm s}$ from $-3$~V to $+3$~V to measure the $\Delta f$-$V_{\rm s}$ curves. (b) $I$-$V$ curves obtained by simultaneously capturing the current between the probe tip and the sample while measuring the $\Delta f$-$V_{\rm s}$ curve. (c) Band alignment of the sample and probe tip considered in the dark state. (d) SPV induced by light irradiation under the condition of $V_{\rm s} = V_{\rm CPD}$.
}
\end{figure}

From the results above, the band structure of the sample and probe tip, as considered in the dark state, is shown in Fig.~2(c). Bulk WSe$_2$ exhibits $p$-type behavior internally, with an electron affinity of $\sim 4.1$~eV and a band gap of $\sim 1.3$~eV\cite{Yoshida_APEX, Smyth}. Near the surface, theoretical calculations and results of photoelectron spectroscopy have shown the presence of inherent surface band bending\cite{Liu}. Furthermore, Se vacancy defects on the surface have also been reported to generate donor-like states\cite{Kim}. Consequently, as shown in Fig.~2(d), a very large SPV is generated, which is the origin of the $V_{\rm cpd}$ variation shown in Fig.~ 2(a).

Next, we present the results of time-resolved measurements conducted at three different light intensities in Figs.~3(a)-3(c). The measurements were conducted under stable feedback conditions in the repulsive regime ($\Delta f = +1$~Hz, corresponding tip-sample force $\sim 10$~nN in this setup), with an amplitude of 4~nm, the $V_{\rm s}$ of approximately 0~V, and the $I_{\rm t}$ of less than 10~fA. The mid light intensity corresponds to that used for the $V_{\rm cpd} = 0.92$~V condition in Fig.~2(a). We successfully obtained a time-resolved signal with a very high S/N ratio as shown in Fig.~3(a). Fitting the data with a two-component exponential function ($A_{\rm fast} \exp (-t/\tau_{\rm fast})+ A_{\rm slow}\exp (-t/\tau_{\rm slow})$) yielded $\tau_{\rm fast} \sim 30$~ns and $\tau_{\rm slow} \sim 150$~ns (Fig.~ 3(b)). 

A schematic model explaining the observed results is shown in Figs.~3(d)-3(g). Previously, when measured by time-resolved STM, both fast and slow components were identified\cite{Yoshida_APEX}. The slow component on the 100~ns scale was considered to correspond to carrier diffusion in the bulk, because the carrier lifetime inside the crystal is extremely long on the $\mu$s scale. On the other hand, the fast component was considered to correspond to the process of extracting electrons that accumulated on the surface as the tunneling current (not shown here). Indeed, for a current of 500~pA, a lifetime of $\tau_{\rm fast} \sim 4$~ns was observed. As the current decreases (i.e., increasing the distance between the probe and the sample, thus extracting fewer electrons), the lifetime increases (250~pA leads to $\tau_{\rm fast} \sim 8$~ns), providing a good explanation for the experimental findings. 

The slow component in this study roughly matches the results obtained from STM measurements, and this component can be attributed to the diffusion process within the bulk (Fig.~3(g)). The fast component is related to the relaxation of photoexcited carriers on the surface (initial relaxation of SPV, Fig.~3(f)), similar to the case with time-resolved STM. However, the fast component in this experiment is about $\sim 30$~ns, slightly longer than that in STM measurements. We conducted experiments using the AFM system as a probe to exploit the force interactions under conditions where almost no tunneling current flows ($V_{\rm s} \sim 0$~V, $I_{\rm t} < 10$~fA). Therefore, the relaxation of about $\sim 30$~ns after photoexcitation can be attributed to the recombination or capture of surface electrons via the gap states suggested by the $I$-$V$ curve measurements. This process is considered the reason why the lifetime appears slightly longer than that in the STM experiment. 

For the data obtained at the three excitation intensities, the results of fitting analysis using a two-component exponential function are shown in Figs.~3(b) and 3(c). These results are derived by averaging the fitting outcomes on both the positive and negative sides of the delay time. As shown in Fig.~3(b), the time constant did not change significantly with intensity, but the signal strength shown in Fig.~3(c) changed. Concerning the signal strength, when the light intensity was reduced (0.6~mW), the intensity ratio of the fast component to the slow one became particularly large ($A_{\rm fast}/A_{\rm slow} \sim 2.14$ (0.6~mW), $\sim 0.89$ (1.1~mW), and $\sim 1.13$ (2.3~mW)). This trend also matches the results of time-resolved STM measurement. In the case of STM, under the weak excitation condition, a small amount of photoexcited carriers were easily pulled out from the surface as tunneling current; therefore, a less diffusion component from the bulk was observed. On the other hand, as the excitation became stronger, a slow component became evident owing to the bulk-side carriers diffusing in the direction perpendicular to the surface after the faster recombination processes of photoexcited carriers via tunneling. The observation that the time constant did not change with light intensity in this study also supports the idea that the recombination processes via the surface states determine the decay rate. The time-resolved signals obtained in this study are consistent with the results obtained by time-resolved STM, indicating the successful measurement of the dynamics of photoexcited carriers using the system developed.

\begin{figure}
\includegraphics[width=8cm]{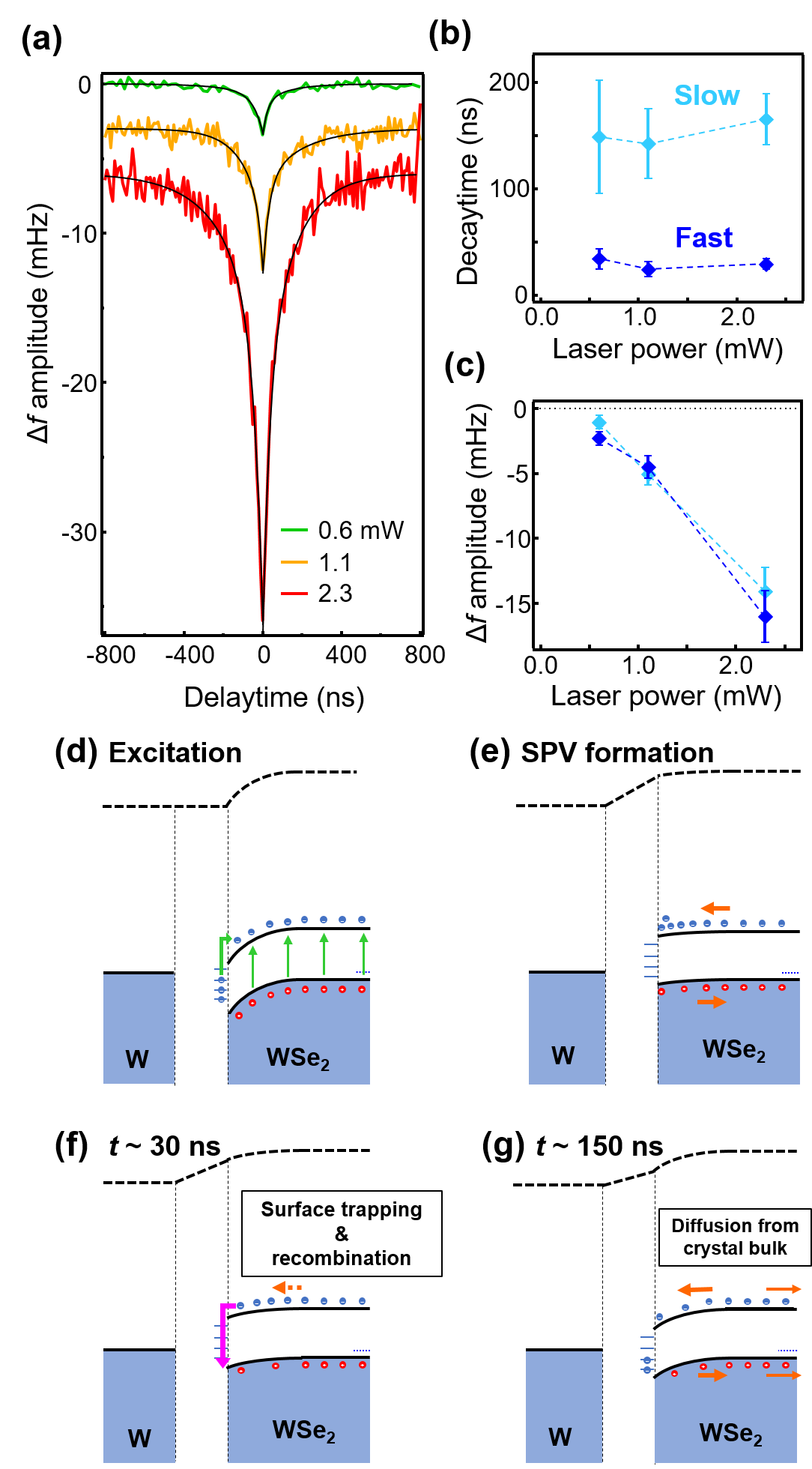}
\caption{
(a) Time-resolved force signals obtained for three different excitation light intensities. Decay time (b) and $\Delta f$ amplitude (c) obtained from fitting the results shown in (a). (d)-(g) Model explaining the results obtained by time-resolved force microscopy.
}
\end{figure}

Owing to the use of delay time modulation, the typical forces that the probe experiences when light is irradiated onto a semiconductor sample are the force induced by SPV (Fig.~4(a)) and the image dipole force discussed for PiFM (Fig.~4(b)). The SPV-induced force is a coulomb force acting between the probe tip and the sample due to the photoexcited carriers that accumulate under the probe after photoexcitation (electrons in this case, Fig.~2(d)). Under the experimental conditions where the distance between the probe tip and the sample does not change with light irradiation (slow feedback condition), the change in SPV due to light irradiation corresponds to the change in $V_{\rm CPD}$, producing the force between the probe tip and the sample. As long as SPV exists under the probe tip, the force due to it will persist, and its time integral becomes the measured force signal. 

On the other hand, according to the discussion on PiFM, the gradient and scattering forces operate owing to the interaction with the optical field\cite{Jahng_PRB}. When a dipole is induced on the probe tip side by the optical field, an image dipole is induced below the probe tip by assuming a semi-infinite sample shape as the experiments in Fig.~3(a)-(c). As a result, an attractive force $F_{\rm image} \propto {\rm Re}\{\alpha_{\rm t}^* \alpha_{\rm s}\} |E_{\rm 0} |^2$ operates between these dipoles\cite{Jahng, Jahng_PRB}. Here, $\alpha_{\rm t}$ and $\alpha_{\rm s}$ are the complex effective polarizabilities of the tip and sample, respectively, within the point dipole approximation. $E_{\rm 0}$ is the incident light electric field vectors. When using pulsed light, a dipole is momentarily induced within the pulse width, and this force operates while plasmons exist on the probe tip side. The lifetime of plasmons in metals is very short on the fs scale compared with that within the laser pulse width (45~ps) used in this study\cite{Brongersma}, so it can be approximated that the force which reflects instantaneous $\alpha_{\rm s}$ only operates while the laser pulse is being applied. 

Figures~4(c) and 4(d) show how the SPV-induced or image dipole force is detected in time-resolved measurement, respectively. Considering the ultrafast time domain, as shown in Figs.~3(d) to 3(e), there is a certain delay from when the carriers are excited until they accumulate on the surface, leading to corresponding change in the generating attractive interactions (rising time shown in Figs.~4(c) and 4(d), see Note 2 in Supplementary Information for more details). When the delay time $t_{\rm d}$ is sufficiently long, the responses to the pump and probe light excitations are induced equally. Let us consider the case of irradiating the probe pulse while photoexcited carriers by the pump pulse are present by reducing $t_{\rm d}$. For simplicity, assuming the flat band condition (where the amount of photoexcited carriers on the surface is saturated, therefore SPV and $\alpha_{\rm s}$ also saturate for each light irradiation), the change induced by the probe light will increase to the same peak height as that caused by the pump light. 

For the case of the force induced by SPV, as previously mentioned, since the force induced by SPV appears as a time integral (green area in Fig.~4(c)), the shorter the delay time, the weaker the measured attractive force becomes. This trend is similar to the signal detection principle of time-resolved STM when optical absorption bleaching is present\cite{Terada, Iwaya, Mogi_APEX1}. 

On the other hand, for the image dipole force shown in Fig.~4(b), when contemplating the force transient response in the high-speed regime, the force is only measured during light irradiation that is sufficiently shorter than the rising time of $\alpha_{\rm s}$ due to the carrier accumulation on the surface, as shown by the orange and red bars in Fig.~4(d). Furthermore, when using delay time modulation, since the value at longer delay times serves as a reference, the change in the magnitude of the force induced by the pump light indicated by the red bars in Fig.~4(d) is probed at $t_{\rm d}$, $\Delta F_{\rm image} = F_{\rm image}(t_{\rm d})$ $-$ $F_{\rm image}(t_{\rm d, ref})$. Therefore, a larger attractive force is detected for shorter delay times. That is, the observed signal with the delay time modulation shows opposite polarity in case of the force derived from SPV shown in Fig.~4(c) and image dipole shown in Fig.~4(d).

\begin{figure}
\includegraphics[width=10cm]{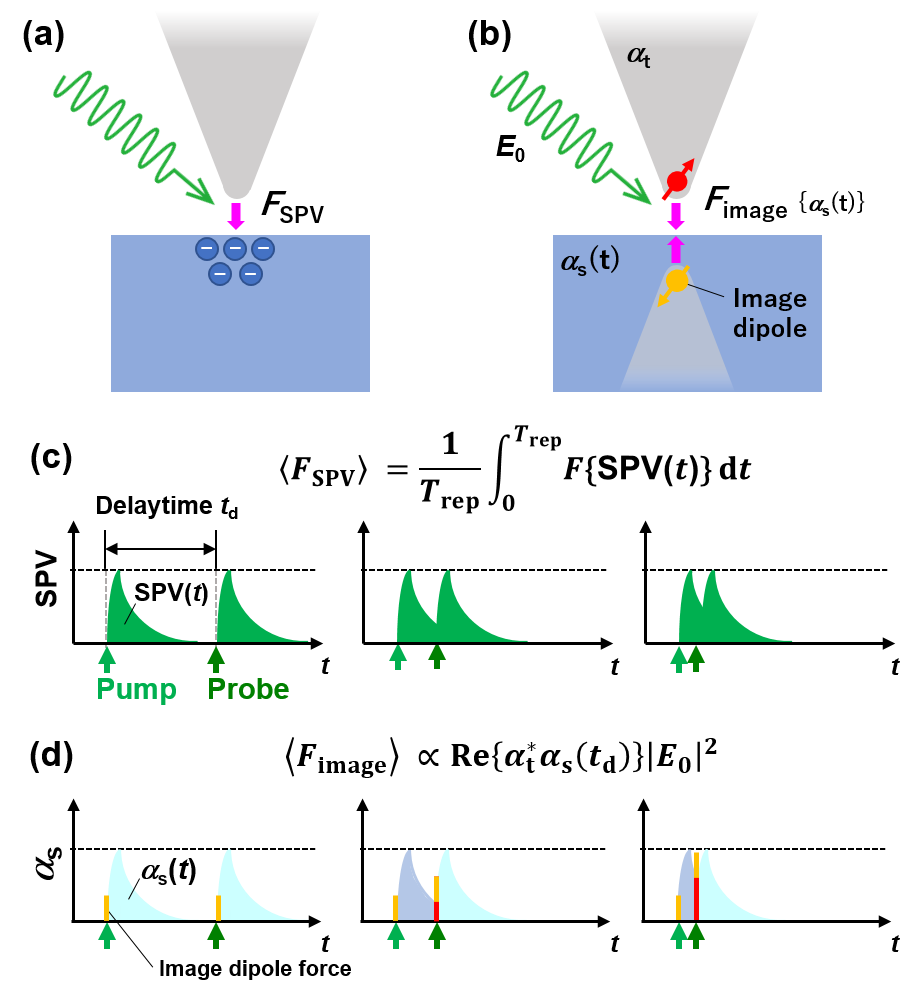}
\caption{
Schematics explaining the observed attractive forces through (a), (c) formation of SPV and (b), (d) dipole-dipole interaction mechanisms. $\alpha_{\rm t}$ and $\alpha_{\rm s}$ are the complex effective polarizabilities of the tip and sample, respectively, within the point dipole approximation. $E_{\rm 0}$ is the incident light electric field vectors. <$F_{\rm SPV}$> and <$F_{\rm image}$> represent time averaged force owing to surface photovoltage and image dipole forces, respectively. $T_{\rm rep}$ is repetition period of the pump-probe laser pulses. In figure (d), the instantaneous image dipole forces are shown as height of orange and red bars, and differences from the largest delay time condition (left figure) are indicated as red.
}
\end{figure}

Here, the latter mechanism corresponds well to the experimental results. To clarify this point, we measured $\Delta f$ responsive to the delay time modulation and the time variation of the phase (Fig.~S1). The attractive force was stronger when the delay was shorter, indicating that we captured the time-resolved signal through the image dipole force. By utilizing the dipole that forms momentarily when the probe light is irradiated, we successfully measured the carrier dynamics of the sample excited by the pump light, such as recombination through intragap states and diffusion. Using a tuning fork and setting the measurement conditions to the repulsive region with a small amplitude of 4~nm, it is considered that the measurement of the near-field forces by PiFM became straightforward.

In conclusion, we developed a time-resolved force microscopy technique by integrating AFM using a tuning fork-type cantilever with the delay time modulation method for optical pump-probe light. We successfully measured the dynamics of surface recombination and diffusion of photoexcited carriers in bulk WSe$_2$, which is challenging owing to the effect of tunneling current in time-resolved STM. By combining with multiprobe systems and using even shorter-pulse laser light\cite{Mogi_NPJ, Mogi_APEX2}, we anticipate new developments such as time-resolved measurements of light-induced ultrafast non-equilibrium dynamics, including measurements in the operation conditions of nanostructures and device structures.

\begin{acknowledgments}
We acknowledge the financial supports of a Grant-in-Aid for Scientific Research (20H00341, 22H00289, 22K14597, 23H00264,) from Japan Society for the Promotion of Science, Japan Core Research Evolutional Science and Technology (CREST) (JPMJCR1875), and JST PRESTO (JPMJPR22AA).
\end{acknowledgments}

\section*{Data Availability Statement}
The authors declare no competing interests.

\section*{Supporting Information}
The Supporting Information is available at \href{https://iopscience.iop.org/article/10.35848/1882-0786/ad0c04}{DOI 10.35848/1882-0786/ad0c04}.


\begin{thebibliography}{99}
\bibitem{Terada} Y. Terada, S. Yoshida, O. Takeuchi and H. Shigekawa, \href{https://www.nature.com/articles/nphoton.2010.235}{Nat. Photonics \textbf{4}, 869 (2010).}
\bibitem{Yoshida_NatNanotech} S. Yoshida, Y. Aizawa, Z.-H. Wang, R. Oshima, Y. Mera, E. Matsuyama, H. Oigawa, O. Takeuchi and H. Shigekawa, \href{https://www.nature.com/articles/nnano.2014.125}{Nat. Nanotechnol. \textbf{9}, 588 (2014).}
\bibitem{Yoshida_ACSPhoton} S. Yoshida, Y. Arashida, H. Hirori, T. Tachizaki, A. Taninaka, H. Ueno, O. Takeuchi and H. Shigekawa, \href{https://pubs.acs.org/doi/10.1021/acsphotonics.0c01572}{ACS Photonics \textbf{8}, 315 (2021).}
\bibitem{Mogi_NPJ} H. Mogi, Y. Arashida, R. Kikuchi, R. Mizuno, J. Wakabayashi, N. Wada, Y. Miyata, A. Taninaka, S. Yoshida, O. Takeuchi and H. Shigekawa, \href{https://www.nature.com/articles/s41699-022-00345-1}{npj 2D Mater. Appl. \textbf{6}, 72 (2022).}
\bibitem{Arashida_ACSPhoton} Y. Arashida, H. Mogi, M. Ishikawa, I. Igarashi, A. Hatanaka, N. Umeda, J. Peng, S. Yoshida, O. Takeuchi and H. Shigekawa, \href{https://pubs.acs.org/doi/10.1021/acsphotonics.2c00995}{ACS Photonics \textbf{9}, 3156 (2022).}
\bibitem{Wang} L. Wang, Y. Xia and W. Ho, \href{https://www.science.org/doi/10.1126/science.abn9220}{Science \textbf{376}, 401 (2022).}
\bibitem{Plankl} M. Plankl, P. E. Faria Junior, F. Mooshammer, T. Siday, M. Zizlsperger, F. Sandner, F. Schiegl, S. Maier, M. A. Huber, M. Gmitra, J. Fabian, J. L. Boland, T. L. Cocker and R. Huber, \href{https://www.nature.com/articles/s41566-021-00813-y}{Nat. Photonics \textbf{15}, 594 (2021).}
\bibitem{Jelic} V. Jelic, K. Iwaszczuk, P. H. Nguyen, C. Rathje, G. J. Hornig, H. M. Sharum, J. R. Hoffman, M. R. Freeman and F. A. Hegmann, \href{https://www.nature.com/articles/nphys4047}{Nat. Phys. \textbf{13}, 591 (2017).}
\bibitem{Yoshioka} K. Yoshioka, I. Katayama, Y. Minami, M. Kitajima, S. Yoshida, H. Shigekawa and J. Takeda, \href{https://www.nature.com/articles/nphoton.2016.205}{Nat. Photonics \textbf{10}, 762 (2016).}
\bibitem{Abdo} M. Abdo, S. Sheng, S. Rolf-Pissarczyk, L. Arnhold, J. A. J. Burgess, M. Isobe, L. Malavolti and S. Loth, \href{https://pubs.acs.org/doi/10.1021/acsphotonics.0c01652}{ACS Photonics \textbf{8}, 702 (2021).}
\bibitem{Hamers} R. J. Hamers and D. G. Cahill, \href{https://pubs.aip.org/aip/apl/article-abstract/57/19/2031/525027/Ultrafast-time-resolution-in-scanned-probe?redirectedFrom=fulltext}{Appl. Phys. Lett. \textbf{57}, 2031 (1990).}
\bibitem{Jahng_APL} J. Jahng, J. Brocious, D. A. Fishman, S. Yampolsky, D. Nowak, F. Huang, V. A. Apkarian, H. K. Wickramasinghe and E. O. Potma, \href{https://pubs.aip.org/aip/apl/article/106/8/083113/1078090/Ultrafast-pump-probe-force-microscopy-with}{Appl. Phys. Lett. \textbf{106}, 083113 (2015).}
\bibitem{Schumacher_PNAS} Z. Schumacher, R. Rejali, R. Pachlatko, A. Spielhofer, P. Nagler, Y. Miyahara, D. G. Cooke and P. Grütter, \href{https://www.pnas.org/doi/full/10.1073/pnas.2003945117}{Proc. Natl. Acad. Sci. U. S. A. \textbf{117}, 19773 (2020).}
\bibitem{Araki} K. Araki, Y. Ie, Y. Aso, H. Ohoyama and T. Matsumoto, \href{https://www.nature.com/articles/s42005-019-0108-x}{Commun. Phys. \textbf{2}, 10 (2019).}
\bibitem{Schumacher_APL} Z. Schumacher, A. Spielhofer, Y. Miyahara and P. Grutter, \href{https://pubs.aip.org/aip/apl/article-abstract/110/5/053111/33401/The-limit-of-time-resolution-in-frequency?redirectedFrom=fulltext}{Appl. Phys. Lett. \textbf{110}, 053111 (2017).}
\bibitem{Kuroiwa} T. Kuroiwa and T. Takahashi, \href{https://iopscience.iop.org/article/10.35848/1347-4065/ac5fbb}{Jpn. J. Appl. Phys. \textbf{61}, SL1004 (2022).}
\bibitem{Rajapaksa} I. Rajapaksa, K. Uenal and H. K. Wickramasinghe, \href{https://pubs.aip.org/aip/apl/article-abstract/97/7/073121/339715/Image-force-microscopy-of-molecular-resonance-A?redirectedFrom=fulltext}{Appl. Phys. Lett. \textbf{97}, 3 (2010).}
\bibitem{Yamanishi} J. Yamanishi, H. Yamane, Y. Naitoh, Y. J. Li, N. Yokoshi, T. Kameyama, S. Koyama, T. Torimoto, H. Ishihara and Y. Sugawara, \href{https://www.nature.com/articles/s41467-021-24136-2}{Nat. Commun. \textbf{12}, 3865 (2021).}
\bibitem{Jahng} J. Jahng, H. Kwon and E. Lee, \href{https://www.mdpi.com/1424-8220/19/7/1530}{Sensors \textbf{19}, 1530 (2019).}
\bibitem{Iwaya} K. Iwaya, M. Yokota, H. Hanada, H. Mogi, S. Yoshida, O. Takeuchi, Y. Miyatake and H. Shigekawa, \href{https://www.nature.com/articles/s41598-023-27383-z}{Sci. Rep. \textbf{13}, 818 (2023).}
\bibitem{Mogi_APEX1} H. Mogi, Z. Wang, R. Kikuchi, C. Hyun Yoon, S. Yoshida, O. Takeuchi and H. Shigekawa, \href{https://iopscience.iop.org/article/10.7567/1882-0786/aaf8b2}{Appl. Phys. Express \textbf{12}, 025005 (2019).}
\bibitem{Yoshida_APEX} S. Yoshida, Y. Terada, M. Yokota, O. Takeuchi, Y. Mera and H. Shigekawa, \href{https://iopscience.iop.org/article/10.7567/APEX.6.016601}{Appl. Phys. Express \textbf{6}, 016601 (2013).}
\bibitem{Arai} T. Arai and M. Tomitori, \href{https://journals.aps.org/prl/abstract/10.1103/PhysRevLett.93.256101}{Phys. Rev. Lett. \textbf{93}, 256101 (2004).}
\bibitem{Addou} R. Addou and R. M. Wallace, \href{https://pubs.acs.org/doi/10.1021/acsami.6b08847}{ACS Appl. Mater. Interfaces \textbf{8}, 26400 (2016).}
\bibitem{Smyth} C. M. Smyth, R. Addou, S. McDonnell, C. L. Hinkle and R. M. Wallace, \href{https://iopscience.iop.org/article/10.1088/2053-1583/aa6bea}{2D Mater. \textbf{4}, 025084 (2017).}
\bibitem{Liu} R.-Y. Liu, K. Ozawa, N. Terashima, Y. Natsui, B. Feng, S. Ito, W.-C. Chen, C.-M. Cheng, S. Yamamoto, H. Kato, T.-C. Chiang and I. Matsuda, \href{https://pubs.aip.org/aip/apl/article/112/21/211603/35265/Controlling-the-surface-photovoltage-on-WSe2-by}{Appl. Phys. Lett. \textbf{112}, 211603 (2018).}
\bibitem{Kim} J. Kim, H. Park, S. Yoo, Y. Im, K. Kang and J. Kim, \href{https://onlinelibrary.wiley.com/doi/10.1002/admi.202100718}{Adv. Mater. Interfaces \textbf{8}, 2100718 (2021).}
\bibitem{Jahng_PRB} J. Jahng, J. Brocious, D. A. Fishman, F. Huang, X. Li, V. A. Tamma, H. K. Wickramasinghe and E. O. Potma, \href{https://journals.aps.org/prb/abstract/10.1103/PhysRevB.90.155417}{Phys. Rev. B \textbf{90}, 155417 (2014).}
\bibitem{Brongersma} M. L. Brongersma, N. J. Halas and P. Nordlander, \href{https://www.nature.com/articles/nnano.2014.311}{Nat. Nanotechnol. \textbf{10}, 25 (2015).}
\bibitem{Mogi_APEX2} H. Mogi, Z. Wang, T. Bamba, Y. Takaguchi, T. Endo, S. Yoshida, A. Taninaka, H. Oigawa, Y. Miyata, O. Takeuchi and H. Shigekawa, \href{https://iopscience.iop.org/article/10.7567/1882-0786/ab09b9}{Appl. Phys. Express \textbf{12}, 045002 (2019).}
\end{thebibliography}

\end{document}